\documentclass[10pt,conference]{IEEEtran}
\IEEEoverridecommandlockouts
\usepackage{cite}
\usepackage{amsmath,amssymb,amsfonts}
\usepackage{algorithmic}
\usepackage{graphicx}
\usepackage{textcomp}
\def\BibTeX{{\rm B\kern-.05em{\sc i\kern-.025em b}\kern-.08em
    T\kern-.1667em\lower.7ex\hbox{E}\kern-.125emX}}

\usepackage{enumitem} % for styling lists
\usepackage{siunitx}  % for \num{}
\usepackage{xspace}
\usepackage{caption}
\usepackage{float}
\usepackage{listings}
\usepackage[dvipsnames]{xcolor} % enables ForestGreen, OliveGreen, etc.
\usepackage{url}                % For handling URLs in bibliography
\usepackage{hyperref}           % For clickable references
\usepackage{cleveref}
\usepackage{booktabs}
\usepackage[most]{tcolorbox}
\usepackage{array}
\usepackage{pifont} % for tick and cross
\usepackage{multirow}
\usepackage[numbers,sort&compress]{natbib}
\usepackage{colortbl}
\usepackage{transparent}

\definecolor{mypurple}{HTML}{E1D5E7}
\colorlet{mypurpleopaque}{mypurple!50!white}
\definecolor{mygrey}{HTML}{E0F2F7} % airy cyan

\newcolumntype{C}{>{\centering\arraybackslash}p{0.2em}} % Narrow centered column

\lstdefinestyle{defaultcode}{
    language=Python,
    basicstyle=\ttfamily\scriptsize,
    keywordstyle=\color{blue},
    stringstyle=\color{red},
    commentstyle=\color{gray},
    showstringspaces=false,
    frame=single,
    breaklines=true,
    morekeywords={assert}
}

\lstdefinestyle{promptstyle}{
  basicstyle=\ttfamily\scriptsize,
  backgroundcolor=\color{gray!10}, % soft gray background
  frame=single,
  rulecolor=\color{black!30},
  columns=flexible,
  keepspaces=true,
  moredelim=**[s][\color{blue!100!black}]{<}{>} % highlight <...>
}

\hypersetup{
    colorlinks=true,
    linkcolor=blue,
    citecolor=blue,
    filecolor=magenta,
    urlcolor=blue
}

\newcommand{\ourtool}{\textsc{blast}\xspace}
\newcommand{\brtlong}{issue-reproducing test\xspace}
\newcommand{\brt}{issue-reproducing test\xspace}
\newcommand{\brts}{issue-reproducing tests\xspace}

\newcommand{\gpt}{gpt-4o\xspace}
\newcommand{\llama}{llama3.3\xspace}
\newcommand{\deepseek}{DeepSeek\xspace}

\newcommand{\N}{\num{426}\xspace}
\newcommand{\best}{\num{151}\xspace}
\newcommand{\bestpct}{\num{35.4}\%\xspace}

\newcommand{\tddbvlong}{TDD-Bench-Verified\xspace}
\newcommand{\tddbv}{TDD-BV\xspace}
\newcommand{\swebench}{SWE-Bench\xspace}
\newcommand{\sbstbench}{PyngBench\xspace}

\begin{document}

\title{Automated Generation of Issue-Reproducing Tests by Combining LLMs and Search-Based Testing
}

\author{\IEEEauthorblockN{Konstantinos Kitsios}
\IEEEauthorblockA{\textit{Department of Informatics} \\
\textit{University of Zurich}\\
Zurich, Switzerland \\
konstantinos.kitsios@uzh.ch}
\and
\IEEEauthorblockN{Marco Castelluccio}
\IEEEauthorblockA{\textit{Mozilla Corporation} \\
London, UK \\
mcastelluccio@mozilla.com}
\and
\IEEEauthorblockN{Alberto Bacchelli}
\IEEEauthorblockA{\textit{Department of Informatics} \\
\textit{University of Zurich}\\
Zurich, Switzerland \\
bacchelli@ifi.uzh.ch}
}

\maketitle

\begin{abstract}
Issue-reproducing tests fail on buggy code and pass once a patch is applied, thus increasing developers' confidence that the issue has been resolved and will not be re-introduced. However, past research has shown that developers often commit patches without such tests, making the automated generation of issue-reproducing tests an area of interest.
We propose \ourtool, a tool for automatically generating \brts from issue-patch pairs by combining LLMs and search-based software testing (SBST).
For the LLM part, we complement the issue description and the patch by extracting relevant context through git history analysis, static analysis, and SBST-generated tests.
For the SBST part, we adapt SBST for generating \brts; the issue description and the patch are fed into the SBST optimization through an intermediate LLM-generated seed, which we deserialize into SBST-compatible form.
\ourtool successfully generates \brts for
151/426 (35.4\%) of the issues from a curated Python benchmark, outperforming the state-of-the-art (23.5\%).

Additionally, to measure the real-world impact of \ourtool, we built a GitHub bot that runs
\ourtool whenever a new pull request (PR) linked to an issue is
opened, and if \ourtool generates an \brt, the bot
proposes it as a comment in the PR. We deployed the bot in three
open-source repositories for three months, gathering data from 32
PRs-issue pairs. \ourtool generated an \brt in 11 of these cases, which we proposed to the developers. By analyzing the developers'
feedback, we discuss challenges and opportunities for researchers
and tool builders.

\noindent \textit{Data and material: \href{https://doi.org/10.5281/zenodo.16949042}{https://doi.org/10.5281/zenodo.16949042}}
\end{abstract}


\begingroup\renewcommand\thefootnote{}\footnote{
This is the author’s version of the paper accepted for publication in the 40th IEEE/ACM International Conference on Automated Software Engineering (ASE 2025). The final version will be available via IEEE Xplore.}
\addtocounter{footnote}{-1}\endgroup

\section{Introduction}\label{sec:intro}
A software issue is typically an online report that describes a bug or a feature request, occasionally including details like stack traces or expected/observed behavior~\cite{bettenburg2008makes}.
In most cases, developers address these issues through a set of code changes (also known as a \emph{patch}).
In this context, an \emph{\brtlong} is a test accompanying the patch that fails on the unpatched code (validating the presence of the issue) and passes on the patched code (validating its resolution)~\cite{libro}. 
An \brtlong increases confidence that the issue (1) can be replicated and (2) will not be reintroduced in the future.

The importance of issue-reproducing tests accompanying a patch is highlighted by previous work~\cite{developer_survey_on_regr_tests,motivation1,motivation2,motivation3}.
For example, \citet{developer_survey_on_regr_tests} surveyed \num{261} developers, asking their view on the statement: ``\textit{when a bug is fixed, it is good to add a test that covers it}.''
Developers agreed with it with a score of 4.4 on a Likert scale~\cite{likertscale} of 5, demonstrating the importance of \brts.\footnote{Reviewers may reject patches until developers add a test: \href{https://github.com/sympy/sympy/pull/13647\#issuecomment-347690202}{sympy\#13647}}

\begin{figure}
    \centering
    \includegraphics[trim=4pt 2pt 3pt 5pt, clip, width=\linewidth]{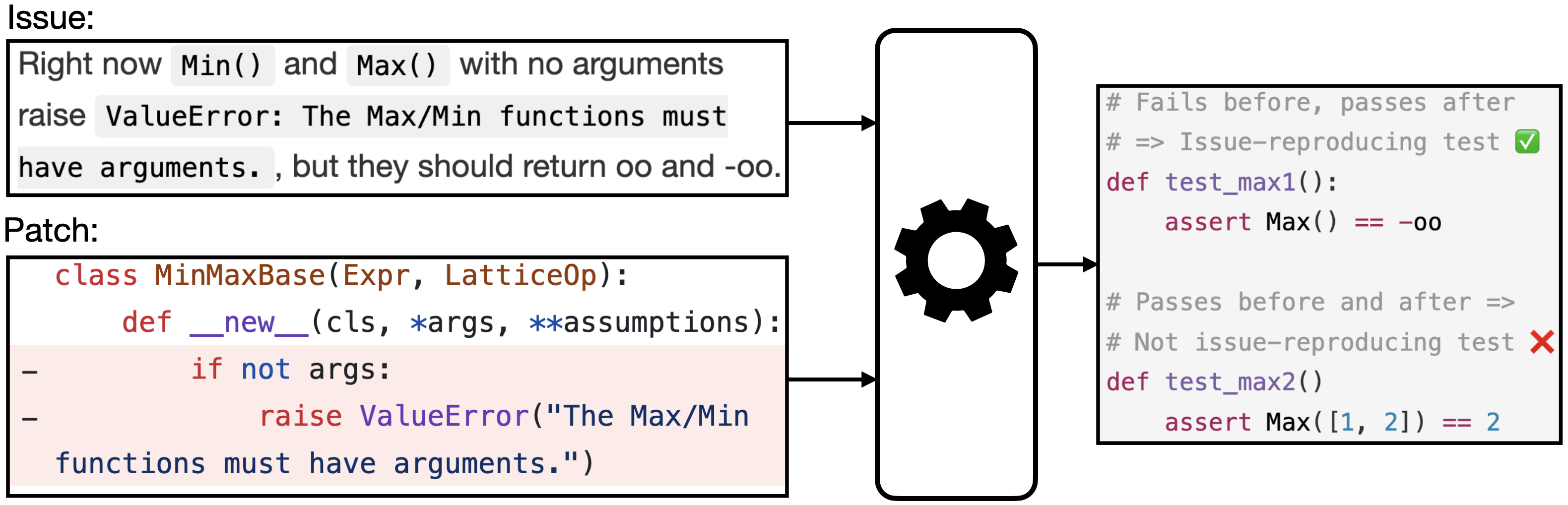}
    \caption{Overview of the task of automatically generating \brts from issue patches.}
    \label{fig:intro}
\end{figure}

Since writing \brts is a demanding and time-consuming task~\cite{timeconsuming1,timeconsuming2} that is often overlooked by developers~\cite{regression_tests_ignored}, recent work has investigated ways to automatically generate these tests to support developers. 
\Cref{fig:intro} schematically presents this automation task.
In particular, recent techniques have proposed the use of Large Language Models (LLMs), either in a zero-shot setting~\cite{tddbench} or in a multi-step workflow where the output of the previous LLM call is used as input to the next one~\cite{tddbench,swtbench,issue2test,otter,fuzzing_regression_tests}.
For example, AutoTDD~\cite{tddbench} feeds the issue description, the patch, and LLM-retrieved code context in an LLM to generate a candidate \brt.

However, relying solely on LLMs for the generation of tests is prone to hallucinations~\cite{hallucinations}, leading for example to tests trying to import non-existent modules or use non-existent methods.
In addition, recent work does not yet leverage information that could be useful for the generation of the \brt such as the focal context or the structure of existing tests for the same class, which could hint the test setup or the available mocks.

To overcome these limitations, we propose \ourtool (generating \textbf{b}ug-reproducing tests using \textbf{L}LMs \textbf{a}nd \textbf{S}BS\textbf{T})~\cite{replication_package}, which uses both LLMs and \emph{search-based software testing} (SBST) for the generation of the tests.
SBST is an approach to test generation~\cite{harman2012search} that combines static and dynamic analysis with genetic algorithms to generate tests for a given module that not only are syntactically correct, but also \textit{pass}.
\ourtool comprises two Components.
The SBST Component generates candidate \brts by running SBST on the patched module.
\ourtool encodes information about the issue into SBST by using an LLM to generate a seed test, and then deserializes it into SBST-compatible form.
The LLM Component generates a candidate \brt by retrieving focal and test context and uses them to build an LLM prompt.
\ourtool also includes in the prompt the SBST tests generated by its first Component, which are always syntactically correct, passing, and test the changed module.

We evaluate \ourtool against a benchmark~\cite{tddbench} of \N issue-patch pairs mined from twelve popular Python repositories and compare it with a zero-shot LLM baseline and the state-of-the-art (SOTA) multi-step LLM workflow~\cite{tddbench}.
\ourtool generates \brts for \bestpct of the issues, outperforming the SOTA (23.5\%) using two LLM queries plus one SBST generation\footnote{The SBST generation has lower cost since it only requires a CPU and a user-defined time budget that could be less than a minute.} instead of three LLM queries.

Evaluating on historical data allows us to run multiple baselines with various settings in a big sample of \N curated issue-patch pairs, but comes with limitations. 
In our case, evaluating on historical data (1) is prone to data leakage that could lead to LLM memorization~\cite{memorization,llm_memorization} and (2) does not allow us to collect the perception of developers on the generated tests.
Other fields evaluating automated methods, such as pull request (PR) reviewer recommendation~\cite{review_recommender} or code completion~\cite{code_completion}, employ a wider range of metrics beyond measuring accuracy on historical data to ensure a more comprehensive view of the performance of a system.
For example, it was found that models for recommending reviewers can achieve up to 92\% accuracy when evaluated against historical data~\cite{reviewer_recom}, but are not considered useful when deployed with developers~\cite{review_recommender}.
Therefore, we develop and deploy a GitHub bot that runs \ourtool whenever a new pull request linked to an issue is opened. 
When \ourtool can generate an \brt, the bot proposes it to the developers by leaving a comment in the PR. 
We deployed the bot to three open-source software repositories for a duration of three months. The bot was triggered in \num{32} PRs linked to an issue and \ourtool generated an \brt in \num{11}/\num{32} cases.
The developers found the tests valid in 6/11 cases---we discuss their feedback to inform future research and practice.

Our work led to the following main research contributions:
\begin{itemize}
    \item \ourtool, a novel technique combining LLMs with SBST for generating \brts;
    \item a dataset for evaluating work on SBST for generating \brts, and a semi-manually filtered, higher-quality version of TDD-Bench-Verified;
    \item evidence of previously unexamined shortcomings of the widely used fail-to-pass metric;
    \item insights for researchers and practitioners from the first evaluation of an \brt generation tool with developers, and the publicly available code of the GitHub bot used in the evaluation.
\end{itemize}

\begin{figure}
\centering
\begin{minipage}{0.48\textwidth}
\begin{lstlisting}[style=defaultcode]
### Module-under-test:
def divide(a: float, b: float) -> float:
    if b == 0:
        raise ValueError("Division by zero")
    return a / b

### Tests generated by Seach-Based Software Testing:
import operations as module_0

def test0():
    float_0 = 10
    float_1 = 2
    assert module_0.divide(float_0, float_1) == 5
\end{lstlisting}
\captionof{figure}{Example of a Pynguin-generated test.}
\label{fig:sbst-tests}
\end{minipage}
\end{figure}

\section{Background}\label{sec:background}
When generating \brts from issue-patch pairs, the \textit{input} consists of (1) an issue description and (2) a patch that resolves the issue (as shown in \Cref{fig:intro}), and the \textit{output} is an \brt.
To generate a test, the most widespread and successful approaches in the literature use either SBST or LLMs~\cite{pynguin,pynguin_time_budget,codamosa,testgen_llm1,issue2test}.
In the following, we describe these two methods, since \ourtool builds on them.

\begin{figure*}
    \centering
    \includegraphics[width=\textwidth]{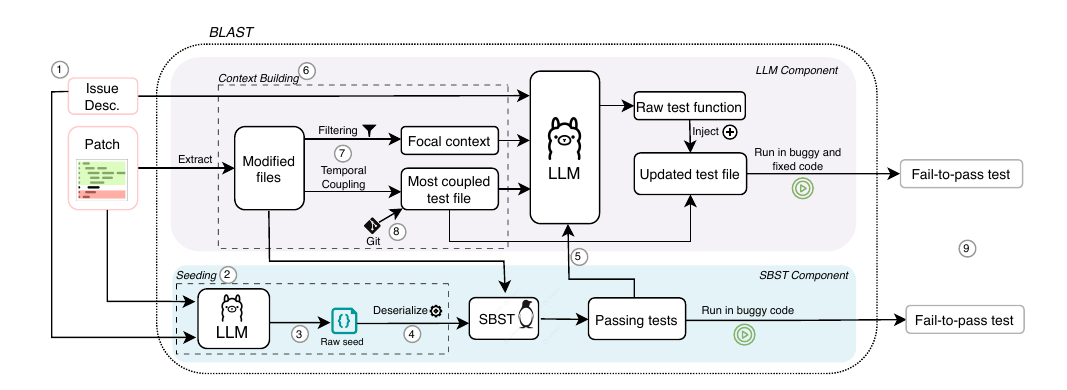}
    \caption{Overview of \ourtool. The \colorbox{mygrey}{SBST Component} and the \colorbox{mypurpleopaque}{LLM Component} coordinate to generate \brts.}
    \label{fig:llm_pipeline}
\end{figure*}

\subsection{Search-Based Software Testing}

The goal of SBST is to automatically generate test cases for an input module-under-test (MUT) by covering diverse behaviors~\cite{evosuite}.
To do so, most SBST tools formulate test generation as an optimization problem:
They start with a randomly generated set of test cases, and mutate them to increase coverage.
Coverage serves as a fitness function that is maximized using meta-heuristic algorithms such as genetic algorithms~\cite{genetic_algorithms} and simulated annealing~\cite{simulated_annealing}.
This optimization continues until the given time budget is exhausted.

Test cases generated by SBST typically consist of sequences of
(i) assignments of variables to random values,
(ii) method or function (callables) calls, and
(iii) assertions that verify the correctness of the call outcomes.  
The callables are identified by analyzing the MUT source- and byte-code through reflection and static analysis. 
The expected values of the assertions are generated based on the \textit{observed} behavior during execution, i.e., SBST generates regression assertions~\cite{regression_assertion}, guarding against future behavior changes. 
For this reason, SBST generates \textit{passing} tests for the MUT.
\Cref{fig:sbst-tests} shows an example of a test generated by Pynguin~\cite{pynguin}, the SOTA SBST tool for Python, for a division function.

\subsection{Large Language Models for Test Generation}
Pretrained LLMs have shown strong capabilities in generating test code from natural language descriptions~\cite{testgen_llm1,testgen_llm2}.
For example, to generate an \brt in the zero-shot setting, it is sufficient to feed to an LLM a prompt that contains the issue description and the patch~\cite{tddbench}.
More recently, workflows consisting of multiple LLM calls have been proposed for code~\cite{autocoderover,repairagent} and test generation~\cite{issue2test,otter,tddbench,fuzzing_regression_tests}.
For example, initial LLM calls are used for action planning~\cite{otter,issue2test} or context retrieval~\cite{tddbench}, and subsequent LLM calls generate and refine the test.

\section{Related Work}\label{sec:related_work}
Previous work has proposed methods for generating \brts from issue-patch pairs.
\textsc{SWT-Agent+}~\cite{swtbench} is a multi-step LLM workflow where in each step the LLM can read/write/edit files or execute a command, and pass the output to the next step. 
AutoTDD~\cite{tddbench} outperformed \textsc{SWT-Agent+} using a three-step LLM workflow to select a test file, retrieve relevant context, and generate the test.
Also, \citet{tddbench} introduce a zero-shot approach with a single LLM query and no context retrieval.
These approaches use solely LLMs for the generation of the test and are evaluated only on historical data, which poses memorization concerns and does not take into account the developer perspective.

Before LLMs, EvoSuiteR~\cite{evosuiter} adapted EvoSuite, the state-of-the-art SBST tool for Java, for differential testing~\cite{differential_testing} between two versions of a codebase.
However, its lack of scalability~\cite{evosuiter}, the unreadable tests~\cite{sbst_readability}, and the lack of encoding of issue information limit its practicality for our task.

Previous work has also explored the adjacent task of generating \brts for test-driven development (TDD) given an issue.
LIBRO~\cite{libro} uses a chain of LLM queries with error feedback to generate \brts in Defects4J~\cite{defects4j}.
Otter~\cite{otter} uses self-reflective action planning with LLMs to plan step-by-step how to generate the test.
Issue2Test~\cite{issue2test} aims at the generation of tests that fail in the unpatched code through three steps of understanding the issue, generating a candidate test, and refining accordingly.
These methods, although effective, focus on the test-driven development scenario of generating an \brt before the developer patch is generated, which is not that widely used in practice~\cite{tdd_in_practice}.

\textsc{Cleverest}~\cite{fuzzing_regression_tests} feeds the commit message and diff to an LLM, prompting it to generate a fail-to-pass input---rather than complete tests---to validate software commits for command-line programs such as XML/PDF parsers.
If the generated input is not already a fail-to-pass input, it is used as a seed for the AFL++ fuzzer~\cite{aflpp}, which mutates it further in an attempt to generate fail-to-pass input.

Finally, Pynguin has been combined with LLMs by CodaMosa~\cite{codamosa} for generating passing tests with high coverage.
When Pynguin reaches a coverage plateau, CodaMosa uses an LLM-generated seed that helps overcome it.
CodaMosa implements a deserialization technique solely based on static rules.
We follow a two-step approach, by accompanying the LLM prompt with a manually curated checklist of what a deserialized test looks like, and then applying static rules to filter out the tests where LLM did not follow our checklist.

\section{\ourtool: Design and Implementation} \label{sec:design}
This section presents the architecture of \ourtool. It takes as input an issue description and the patch resolving that issue (Point 1 in \Cref{fig:llm_pipeline}) and outputs up to two tests that fail in the unpatched code and pass in the patched one (Point 9).
\ourtool consists of an \colorbox{mygrey}{SBST Component} and an \colorbox{mypurpleopaque}{LLM Component}, interacting as shown in \Cref{fig:llm_pipeline}.
The SBST Component builds on Pynguin to generate passing tests for the patched module, by starting the mutations from an LLM-generated (Point 3) and deserialized (Point 4) seed to capture the issue semantics.
The Pynguin-generated tests are then used as both candidate \brts (Point 9) and additional context for the LLM Component (Point 5).
The LLM Component automatically retrieves relevant context (Point 6) from the repository (Point 7), the git history (Point 8), and the Pynguin-generated tests (Point 5), and feeds it to an LLM to generate a second candidate \brt (Point 9).

\subsection{SBST Component}
Traditionally, SBST techniques have been devised, evolved, and evaluated with the goal of generating \textit{passing} tests for a given module that maximize coverage~\cite{pynguin}.
To adapt SBST for generating \brts from issue-patch pairs, \ourtool starts the mutations from an LLM-generated and statically deserialized seed that captures the semantics (e.g., magic values) mentioned in the issue/patch.
Out of the resulting SBST-generated tests, \ourtool extracts those that reveal differences between the patched and unpatched code.
We explain these two mechanisms in more detail below.
Since our work focuses on Python, we will use Pynguin~\cite{pynguin}, the state-of-the-art SBST tool for Python, but \ourtool is extensible to work with EvoSuite~\cite{evosuite} (Java) as well.

\smallskip
\noindent\textbf{Seeding Pynguin.}
Pynguin does not accept natural language input, therefore we cannot steer the test generation towards reproducing a given issue.
Yet, it is possible to start the mutations from a user-provided seed test.

For this to happen, a technical challenge is that the seed must be a test case in \textit{Pynguin-canonical form}, i.e., the internal representation of tests in Pynguin~\cite{pynguin}.
In the absence of related documentation, we analyzed the source code of Pynguin and came up with a checklist of properties acceptable in this form.
The properties vary from simple ones like ``a value should be assigned to a variable before using it as input to a callable'', to more complex ones like only importing callables from modules which the MUT imports.
The full checklist is available in our replication package~\cite{replication_package}.
We prompt an LLM to generate a seed given the issue description, the patch, and our curated checklist.
The LLM generates a compatible seed in 24\% of the cases in our experiments.

Since many seeds are still incompatible, we develop a middleware that statically analyzes the LLM-generated seed to filter out statements that violate the canonical form.
Even if a single statement of the seed violates the form, Pynguin will discard the whole seed, so our middleware focuses on detecting statements that violate the form and discarding them so that the rest of the statements can be used.
Our middleware increases the seed acceptance rate from 24\% to 50\%.
We note that the other 50\% contains cases where a seed cannot be accepted, because to generate a seed related to the issue, the LLM must import code that is not imported in the MUT.
In these cases, \ourtool runs Pynguin starting from random seeds.

\smallskip
\noindent\textbf{Runtime Filtering of Difference-Revealing Tests.}
To generate potentially difference-revealing tests, \ourtool first extracts the module(s) changed by the patch.
Then, it generates tests for \textit{the patched version} of that module(s), which will be, by design, passing.
The objective for Pynguin is the whole patched module, as Pynguin does not support finer-grained objectives like the patched function or the patched lines.
Finally, \ourtool runs the generated tests in the unpatched version, and if a test fails, it is a difference-revealing test, also called \textit{fail-to-pass} (F$\rightarrow$P) test.
Previous research has shown that SBST may generate flaky tests, i.e., tests that either fail or pass on the same code~\cite{sbst_flaky}. 
To account for flakiness, we run the generated test in the patched code as well, and if it fails (which by design should not), we discard it.

The output of \ourtool's SBST Component is (i) a set of tests that pass in the patched code, to be used as prompt context in its LLM Component (Point 5), and (ii) a (potentially empty) set of F$\rightarrow$P tests (Point 9).

\subsection{LLM Component}\label{ssec:design_llm_component}
As mentioned in \Cref{sec:background}, LLMs can generate tests in a zero-shot setting from a prompt containing the issue description and the patch. 
However, this context is often not sufficient to capture the complexity of the task, as hinted by the relatively low number (23.5\%) of \brts generated by the SOTA~\cite{tddbench}.
We identify three additional sources of context: existing tests for the MUT, focal context, and SBST-generated tests that pass in the patched code.

\smallskip
\noindent\textbf{Existing Tests Retrieval.}
Existing tests for the MUT can be useful for two reasons. 
First, they often contain information on how to set up the test or how to mock objects.
Second, the LLM will generate a raw test function, which we must inject into an existing test file.
For these reasons, \ourtool retrieves the test file most coupled to the changes, and uses it to i) feed its existing tests to the LLM, and ii) inject the generated test.

To infer the most coupled test file, \ourtool extracts all the filenames changed in the patch, and applies an initial name-based rule: if \texttt{divide.py} is patched, we search for \texttt{test\_divide.py}.
If no such file exists, which could be the case if the tests lived in \texttt{test\_operators.py}, we propose using \textit{coedit temporal coupling}~\cite{coupling1}, a metric widely used in the mining software repositories community~\cite{coupling2}.
Specifically, \ourtool iterates all the past commits that changed \texttt{divide.py} and searches for the most co-edited file that starts with \texttt{test\_}.
The intuition is that, if whenever \texttt{divide.py} changes, \texttt{test\_operators.py} also changes, then probably the latter contains tests for the former.
In some cases, the retrieved file is very large, which caused the LLM performance to drop and made us feed only the first three tests in the prompt.

Dynamic analysis could also be used to infer the most coupled test file, for example by running  all the tests and selecting the one that covers more lines of the patch. However, this would require significant compute resources and time, hence we opted for the static heuristic described above that requires far fewer resources.

\begin{figure}
\begin{lstlisting}[style=promptstyle]
Below is a software issue: <issue>.

A developer resolved the issue with the following patch:
<patch>.

Additional code context is shown below: <focal_code>.

Existing tests for focal code are shown below:
<retrieved_tests>.

Pynguin-generated tests for the focal code are also shown
below: <pynguin_tests>.

Your task is to generate an issue-reproducing test, i.e., 
a test that fails before and passes after the patch.

\end{lstlisting}
\caption{Outline of the prompt used in the LLM Component of \ourtool. The full prompt is available in our replication package.}\label{fig:prompt}
\end{figure}

\smallskip
\noindent\textbf{Focal Context Retrieval.}
The raw patch contains the lines that changed, which are usually not sufficient for generating an \brt. 
For example, if a method of a class is changed, an \brt would generally contain at least an instantiation of a class object and a call to the changed method.
Hence, the LLM should have information about the class signature, the method signatures, the constructor, and information about other properties/methods of the class that could be used in the test setup.
This context is also referred to as \textit{focal context}~\cite{focal_context}.
\ourtool retrieves the focal context as follows: if the patch changes a class, it retrieves the class' signature, properties, constructor, and methods.
Moreover, \ourtool retrieves all the global statements since they often contain useful configurations. 
If the patch changes a standalone function or expression, \ourtool only retrieves the global statements.

\smallskip
\noindent\textbf{Pynguin-Generated Seeds.}
Finally, \ourtool uses the tests generated by its SBST Component as additional context for the LLM.
The intuition is that syntactically correct context could reduce the LLM hallucinations, as we qualitatively observe in ~\Cref{ssec:results_offline}.
Also, the fact that the tests already pass in the patched code could make the generation of a F$\rightarrow$P test easier.
The number of tests generated by Pynguin is not constant; 
a typical range from our experiments is 1 to 50, depending on the size of the MUT.
To keep the prompt compact, we feed only the first three tests.

\smallskip
\noindent\textbf{Zero-Shot Prompt.}
We construct a zero-shot prompt by providing the issue description, the patch, and the retrieved inputs, and asking the LLM to generate a test function that reproduces the issue by failing before and passing after the patch.
An outline of the prompt is shown in~\Cref{fig:prompt} and is partly inspired by previous work~\cite{swtbench}.
\ourtool feeds the prompt to the LLM, gets back a raw test function, and injects it into the retrieved test as follows:
If the test file contains standalone test functions, we inject the generated test function at the end of the file.
If, however, the test file contains test classes that in turn contain test methods, we inject the generated test function as a method of the last class of the file, after prepending the argument \texttt{self} in the list of arguments.
Previous work has experimented with generating a test patch instead of a raw test~\cite{swtbench}, which would allow editing existing test functions instead of creating new ones and would eliminate the need of injecting the raw test to a file.
However, the LLMs hallucinated the line numbers in the patch, leading to many errors~\cite{tddbench}.
Another alternative is writing each generated test to a new file.
This was not preferred because it reduces the practicality, as the developer would need to move the test to the appropriate file.
Hence, we opted for asking a raw test function and injecting it to the test file retrieved above.

\section{Benchmark-based Evaluation: Methodology}\label{sec:evaluation_offline}
We present here the methodology for the benchmark-based empirical evaluation of \ourtool, which we structure around two research questions. Methodological choices specific to each question are presented in~\Cref{ssec:results_offline}, closer to their results.

\subsection{Research Questions}

First, we compare \ourtool against two baselines on a recently proposed benchmark of \num{449} issue-patch pairs across 12 popular Python repositories~\cite{tddbench}, after applying a semi-manual filtering to discard trivial entries.
\begin{tcolorbox}
\begin{minipage}{\textwidth}
\textbf{RQ1.} How does \ourtool compare to baselines in generating \brts from issue-patch pairs?
\end{minipage}
\end{tcolorbox}

Then, we investigate the contribution of \ourtool's LLM and SBST Components.
For the former, we perform an ablation study of the input combinations across medium- and large-sized LLMs.
For the latter, we study how the LLM seed and the time budget affect the performance and in doing so, we curate and release~\cite{replication_package} the first benchmark for SBST on this task.

\begin{tcolorbox}
\begin{minipage}{\textwidth}
\textbf{RQ2.} How do the components and hyperparameter choices of \ourtool affect its performance?
\end{minipage}
\end{tcolorbox}

\begin{figure}
    \centering
    \includegraphics[width=0.85\linewidth]{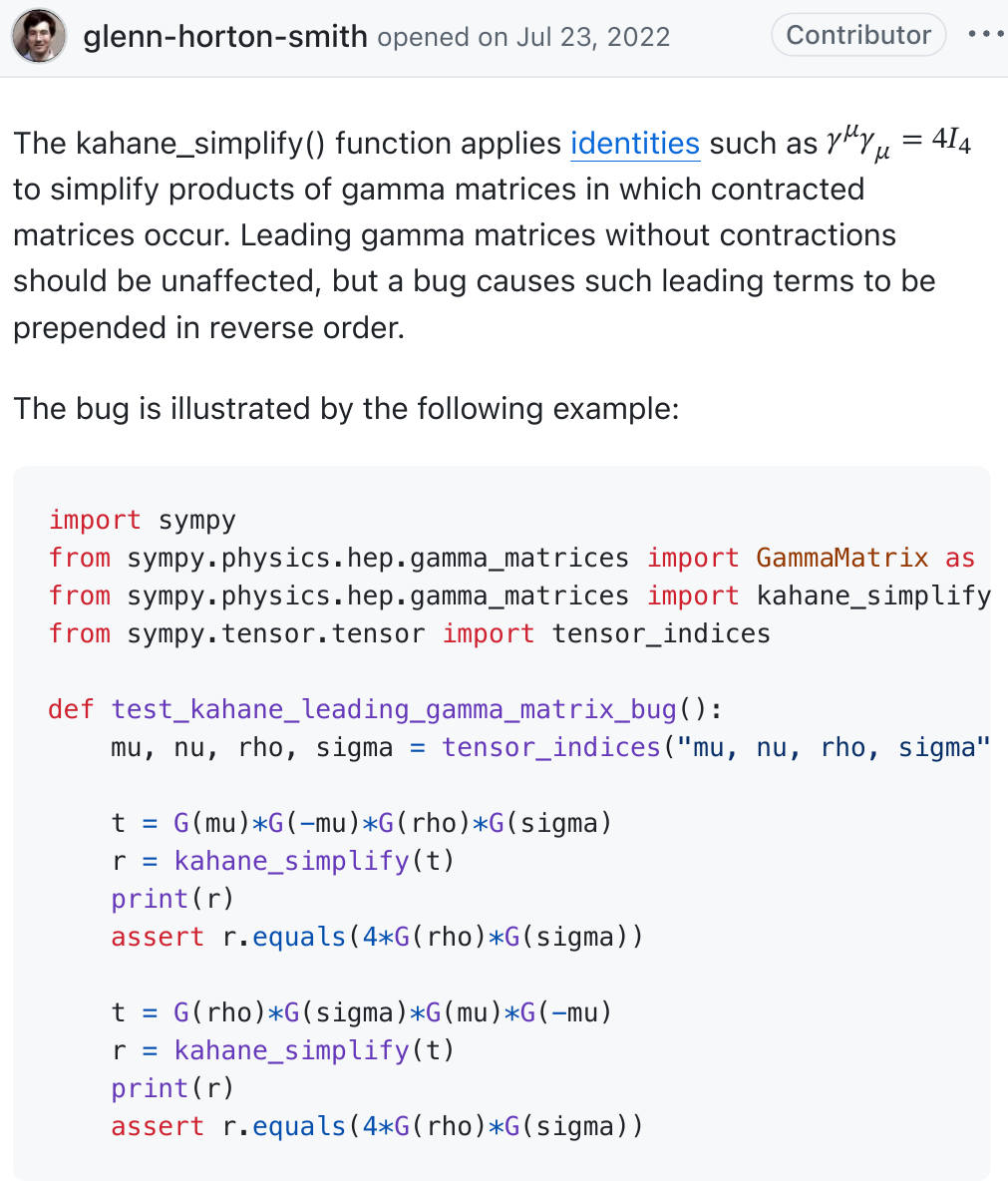}
    \caption{Example of a trivial instance: Given this issue (\href{https://github.com/sympy/sympy/issues/23823}{\#23823} in the \texttt{sympy} repository), generating an \brt is easy for a modern LLM.}
    \label{fig:low_quality}
\end{figure}

\subsection{Datasets}\label{sssec:eva_datasets}
As shown in \Cref{fig:intro}, our inputs are an issue description and a patch for this issue.
We use the \tddbvlong (\tddbv) dataset, which is used in recent literature~\cite{otter,tddbench} and consists of \num{449} issue-patch pairs from \num{12} open-source Python repositories.
Alternative landmark datasets, such as Defects4J~\cite{defects4j}, have been saturated by recent LLMs~\cite{defects4j_saturated}.

\smallskip
\noindent\textbf{Data Cleaning.} \tddbv is an adaptation of the SWE-Bench dataset~\cite{swebench}, which was designed for a different task (i.e., generating a patch given an issue).
This difference led to entries like the one shown in~\Cref{fig:low_quality}:
Although it is acceptable to use this issue as input to generate the patch, generating an \brt given this issue is trivial for modern LLMs, since such a test exists in the input.
To refine \tddbv, we apply a semi-manual filtering process. Specifically, we manually inspect all issues containing the expressions \texttt{def test} or \texttt{assert}, as these suggest the presence of test code. This inspection yields 37 issues, of which 23 are deemed trivial because they already include a complete \brt. In contrast, the remaining issues contain assertions in isolated snippets that merely suggest the test logic, requiring further synthesis by the LLM. We exclude the 23 trivial cases from \tddbv and use the resulting filtered dataset ($N = 426$) in our experiments.

\smallskip
\noindent\textbf{Pynguin-Compatible Dataset.}
SBST tools in the literature~\cite{codamosa,evosuite,harman2012search} are evaluated in benchmarks compatible with SBST requirements, since these tools struggle to analyze modules that rely on heavy introspection or dynamic dispatch~\cite{pynguin}.
Also, only specific versions of Python are supported, namely 3.8--3.10.
In the absence of such a benchmark for our task, we set to construct and release \sbstbench as follows.
We start from the \N entries of \tddbv  and filter out those using unsupported Python versions, resulting in 254 entries.
Then, we set up a Docker container, checkout the patched code, install the appropriate Pynguin version, and run it for one minute to generate tests for the changed modules, following the incompatibility filtering of CodaMosa~\cite{codamosa}.
The generation is successful in 113 cases, fails due to internal errors in 82 cases (e.g., \href{https://github.com/se2p/pynguin/issues/81}{issue \#81}), and fails due to incompatible MUT in 172 cases, leaving \num{113} entries in PyngBench.

We use \sbstbench in~\Cref{ssec:results_rq2} to study the performance of Pynguin in isolation.
In the rest of the paper, \ourtool runs in the whole \tddbv, and when the Pynguin Component fails, \ourtool proceeds with the LLM Component.

\begin{figure}
    \centering
    \includegraphics[trim=27pt 10pt 20pt 10pt, clip, width=0.7\linewidth]{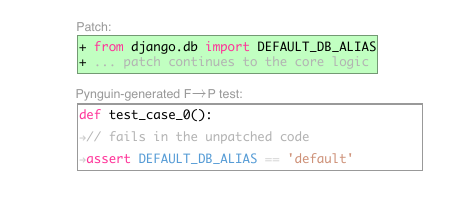}
    \caption{A F$\rightarrow$P test that is not an \brt. 
    }
    \label{fig:f2p_but_not_br}
\end{figure}

\subsection{Evaluation Metric: Fail-to-Pass}

Our goal is to evaluate to what extent \ourtool can generate \brts for a given issue-patch pair.
To operationalize the classification of an automatically generated test as an \brt, previous work~\cite{libro,tddbench,swtbench,issue2test,otter} has adopted the binary metric \emph{Fail-to-Pass} (F$\rightarrow$P). The value of this metric is true \emph{iff} the test fails (F) on the unpatched code and passes (P) on the patched code.
The intuition is that an \brt should expose the issue and prevent a regression (i.e., fail on the unpatched code) as well as confirm the resolution (i.e., pass on the patched code).
In \Cref{fig:intro}, the F$\rightarrow$P metric would be evaluated as true\footnote{Henceforth, we refer to a test for which the \emph{Fail-to-Pass} (F$\rightarrow$P) metric is evaluated to be true as a \emph{Fail-to-Pass test} or \emph{F$\rightarrow$P test}.} for the \brt (above), and as false for the other test (below).

\smallskip
\noindent\textbf{Metric Validation.} While it is reasonable to think that every \brt should be a F$\rightarrow$P test, the inverse may not necessarily hold.
Consider for example the Pynguin-generated test of \Cref{fig:f2p_but_not_br}.
The F$\rightarrow$P metric is true: the \texttt{DEFAULT\_DB\_ALIAS} is not imported in the unpatched code, so the test fails, but it is imported in the patched code, so the test passes. 
However, this test does not reproduce the issue, which is about prefetching sliced queries (\href{https://github.com/django/django/pull/15957/files}{django \#15957}).

For these reasons, we empirically validate---for the first time in the literature---how accurately the results of the F$\rightarrow$P metric can operationalize the classification of \brts.
To this aim, for each F$\rightarrow$P test generated by the Pynguin Component of \ourtool, we manually examine it to judge whether it is a genuine \brt or the metric is evaluated as true for other reasons, such as a side effect in \Cref{fig:f2p_but_not_br}.
When we refer to the number of F$\rightarrow$P tests in the rest of the paper, we mean the manually verified ones for Pynguin, and we use $F \rightarrow P^*$ to indicate all the fail-to-pass tests.
The LLM-generated tests are more in number, so we analyze a sample of \num{50} of them. 
We find that all of them target the core issue, so for LLM-generated tests we assume $F \rightarrow P^* = F \rightarrow P$.
This is because, in contrast to Pynguin, the LLM has explicit access to the issue, which, under our current prompt, makes it generate true \brts.
Given our limited expertise with the 12 repos, we have high confidence for the absence of obvious cases like in \Cref{fig:f2p_but_not_br} but we may have missed borderline cases, which we tackle by evaluating F$\rightarrow$P tests with developers in \Cref{ssec:results_invivo}.

\subsection{Experimental Setup}\label{sssec:experimental_setup}
We ran Pynguin on an Ubuntu 22.04 machine with 16 Intel Xeon CPUs and 128GB RAM, using the random seed \num{42} to increase reproducibility.
We use three LLMs, namely \gpt (gpt-4o-2024-08-06), \llama (llama-3.3-70b-versatile), and \deepseek (deepseek-r1-distill-llama-70b), covering a range of medium- to large-sized models, with and without reasoning.
For \gpt and \llama we use a temperature of \num{0} to increase reproducibility and also follow related work~\cite{tddbench,otter}.
We use the latest version of Pynguin, which is v0.41 at the time of writing.
For instances of \tddbv that run on Python $<$ 3.10 (and are therefore not supported in Pynguin v0.41), we use the most recent version of Pynguin that still supports Python $<$ 3.10 (i.e., Pynguin v0.17). We keep the default Pynguin parameters (e.g., \texttt{DYNAMOSA} algorithm, $crossover\,rate = 0.75$, $population = 50$), unless explicitly specified otherwise in our ablation study (\Cref{ssec:results_rq2}).

\begin{table}
    \centering
    \scriptsize
    \caption{Performance of \ourtool against two baselines.}\label{tab:results_rq1_autotdd}
    \begin{tabular}{llcc}
        %\toprule
        LLM & Method & \# F$\rightarrow$P & \% F$\rightarrow$P \\
        \midrule
        & ZeroShot & 86 & 20.2  \\
        \gpt & AutoTDD  & 100 & 23.5 \\
        & \ourtool (Ours) & \textbf{151} & \textbf{35.4}  \\
        \midrule
        & ZeroShot &   45  & 10.6   \\
        \llama & AutoTDD  & 51 & 12.0   \\
        & \ourtool (Ours) &  \textbf{131} & \textbf{30.8} \\
        \midrule
        & ZeroShot &  43 & 10.1 \\
        \deepseek & AutoTDD  & 48 &  11.3 \\
        & \ourtool (Ours) & \textbf{114} & \textbf{26.8} \\
    \end{tabular}
\end{table}

\begin{table}
\centering
\scriptsize
\caption{Input combinations ($C_1$--$C_7$) and their effect on \ourtool's LLM Component (underlying model in \gpt).}
\label{tab:res_rq1_ablation}
\begin{tabular}{c@{\hskip 0.05in}|c@{\hskip 0.05in}|c@{\hskip 0.05in}|c@{\hskip 0.05in}|c@{\hskip 0.05in}|c@{\hskip 0.05in}||r@{\hskip 0.1in}c@{\hskip 0.05in}c@{\hskip 0.05in}}
%\toprule
\multicolumn{6}{c||}{\textbf{Input Combinations}} & \multicolumn{3}{c}{\textbf{Results} ($\mathbf{F \rightarrow P}$)} \\
\cmidrule{1-6} \cmidrule{7-9} 
\textbf{ID} & \textbf{Issue} & \textbf{Patch} & \textbf{Focal} & \textbf{Ex. Tests} & \textbf{P. Tests} & \textbf{\#} & \textbf{\%} & \textbf{Cum. \%}   \\

\midrule
$C_1$ &\ding{51} & &  & \ding{51} &   & 84 & 19.7 & 19.7  \\
$C_2$ & & \ding{51} & \ding{51} &  & & 86 & 20.2 & 32.6  \\
$C_3$ & \ding{51} & \ding{51} &  & & & 124 & 29.1 & 42.3  \\
$C_4$ & \ding{51} & \ding{51} & \ding{51} &  &  & 137 & 32.2 & 45.1 \\
$C_5$ & \ding{51} & \ding{51} & \ding{51} & \ding{51} &  & 140 & 32.9 & 48.8  \\
\cellcolor{gray!20}$C_6$ & \cellcolor{gray!20} \ding{51} & \cellcolor{gray!20}\ding{51} & \cellcolor{gray!20}\ding{51} & \cellcolor{gray!20}  & \cellcolor{gray!20} \ding{51} & \cellcolor{gray!20} \textbf{145} & \cellcolor{gray!20}\textbf{34.1} & 49.3 \\
$C_7$ & \ding{51} & \ding{51} & \ding{51} & \ding{51} & \ding{51} & 143 & 33.6 & 49.5  \\
\end{tabular}
\end{table}

\section{Benchmark-based Evaluation: Results}\label{ssec:results_offline}
We present further methodological details for each research question and report the corresponding results.

\subsection{RQ1 - Performance}

To demonstrate the effectiveness of \ourtool, we compare it to two baselines from the literature.
AutoTDD~\cite{tddbench} achieves SOTA results in generating \brts from issue-patch pairs using the three-step LLM workflow described in \Cref{sec:related_work}.
ZeroShot is a single-step baseline from Ahmed et al.~\cite{tddbench} that feeds the issue description and the patch to the LLM, without any additional retrieved context.

\Cref{tab:results_rq1_autotdd} presents the results:
\ourtool outperforms previous methods across all underlying LLMs evaluated.
The best performance is with \gpt, where AutoTDD generates \brts in 100/\N (23.5\%) of the issues, while \ourtool does sp for \best/\N (\bestpct).
Regarding overlap, \ourtool generates \brts for 67 issues that AutoTDD misses, while AutoTDD does so for 16 issues that \ourtool misses.

\begin{tcolorbox}
\begin{minipage}{\textwidth}
\textbf{Finding 1.} \ourtool could generate \brts for \best/\N of the cases---an absolute increase of 51 cases over the state-of-the-art (100/426).
\end{minipage}
\end{tcolorbox}

\ourtool differs from AutoTDD in two key aspects: (1) the augmented context provided in the LLM prompt, and (2) the SBST Component. In RQ2, we analyze the contribution of each of these aspects to the performance improvement.

We also report the patch coverage~\cite{patch_cov}, i.e., the patch lines that are covered by the F$\rightarrow$P tests.
The coverage ranges from \num{91}\% for ZeroShot to \num{94}\% for \ourtool but the difference is not statistically significant.
Previous work has also observed that for F$\rightarrow$P tests the coverage is high regardless of the generation method~\cite{swtbench,tddbench}, so the comparison between two methods is reduced to the comparison of their F$\rightarrow$P rate.

Regarding the cost of each approach,
Pynguin achieves the best result with a time budget of \num{60} seconds.
We assume that an LLM query and the context extraction in both \ourtool and AutoTDD have zero time overhead; 
In reality, these processes take up to a few seconds, which is negligible compared to the 60 seconds.
Under these assumptions, \ourtool requires two LLM queries and 60 seconds time overhead, while AutoTDD requires three LLM queries with no time overhead.

Finally, to evaluate how well \ourtool retrieves the most related test file, we compare the retrieved file with the file where the developer placed their test in the actual PR.
\ourtool's name- and git-based heuristic matches the developer file in 86\% of the cases, compared to AutoTDD's LLM-based retrieval which only matches the developer's file in 61\% of the cases.

\subsection{RQ2 - Ablation}\label{ssec:results_rq2}
As shown in \Cref{fig:llm_pipeline}, \ourtool consists of an LLM Component and an SBST Component, each generating a candidate test.
We investigate how the two Components and their hyperparameters contribute to the result of \Cref{tab:results_rq1_autotdd}.

\smallskip
\noindent\textbf{LLM Component.}
To assess the impact of additional prompt information on the performance of the LLM Component, we begin with the baseline prompt described in~\Cref{sec:design} and construct seven variants, each including only a specific subset of the prompt inputs.
\Cref{tab:res_rq1_ablation} details the configurations of these seven input combinations.
The first combination ($C_1$) corresponds to the test-driven development case (as investigated in previous work~\cite{tddbench,swtbench,otter,issue2test,libro}) where the test must be written before the patch and only the issue description is available with potential example tests.
In $C_2$, we only provide the issue title without description to simulate scenarios where the issue description is absent or very poor, as may happen in practice~\cite{bettenburg2008makes}.
$C_3$ uses only the issue description and the patch which are readily available, and we use it to measure the performance increase with our focal code retrieval ($C_4$), test retrieval ($C_5$), and SBST-generated tests ($C_6$, $C_7$).
We run \ourtool with each input combination $C_i$ and report the results for our best model  \gpt in \Cref{tab:res_rq1_ablation}, while the results for the other models exist in our replication package~\cite{replication_package}.

For the TDD scenario ($C_1$), \ourtool generates a F$\rightarrow$P test in 19.7\% of the cases even though it was not designed for TDD.
When only the issue title is given instead of the description ($C_2$), \ourtool performs significantly worse (8.9\%) than when the description is added ($C_3$).
This demonstrates the relevance of a good issue description for test generation tools.
By adding focal context ($C_4$), the performance increases by another 3.1\%.
The retrieved test case ($C_5$) yields small increase (0.7\%) in \gpt, but yields the largest increase for \llama (6.3\%).
Incorporating Pynguin-generated tests into the prompt ($C_6$, $C_7$) yields the best performance (+1.2), even though Pynguin-generated tests were only available for \num{113} out of the \N cases.
\deepseek and \llama follow a similar trend with small differentiations that we analyze in our replication package~\cite{replication_package}, with gpt-4o outperforming them across all $C_i$.

Unexpectedly, the cumulative number of F$\rightarrow$P tests generated by the union of all combinations (last column of \Cref{tab:res_rq1_ablation}) is 15.4\% higher than the best-performing individual run.
We attribute this to the fact that LLMs have been shown to prioritize content towards the end of the prompt, known as recency bias~\cite{recency_bias}. 
So, when we add for example the existing tests at the end of the prompt ($C_5$), the focal code that is earlier in the prompt now, receives less attention.

\begin{table}
\centering
\scriptsize
\caption{Effect of Pynguin variations (N=113).}
\label{tab:swtbench_p}
\begin{tabular}{lr|rc|cc}
\textbf{Variation} & \textbf{\#F$\rightarrow$P\textsuperscript{*}} & \textbf{\#F$\rightarrow$P} & \textbf{\% F$\rightarrow$P} & \textbf{\#F$\rightarrow$P Unq.} & \textbf{\% F$\rightarrow$P Unq.}  \\
\midrule
$t=6$ & 10 & 7 & 6.2 & 4 & 3.5 \\
$\mathbf{t=60}$  & \textbf{13} & \textbf{10}  & \textbf{8.8} & \textbf{6} & \textbf{5.3} \\
$t=600$ & 12 & 9 & 7.9 & 5 & 4.4\\
\midrule
No seed    & 8 & 4 & 3.5 & 3 & 2.7 \\
\end{tabular}
\end{table}

\smallskip
\noindent\textbf{SBST Component.}
We run a set of experiments disabling \ourtool's LLM Component  to measure the performance of the SBST Component and explore the impact of various parameters.
For the methodological reasons explained in \Cref{sssec:eva_datasets}, we run this set of experiments against the \sbstbench dataset.

To analyze the impact of time budget on performance, we run Pynguin with three configurations: the default $t=600$ seconds commonly used in prior studies~\cite{codamosa, pynguin_time_budget}, a practical $t=60$ seconds suitable for real-time GitHub bot usage, and a minimal $t=6$ seconds to approximate the latency of a typical LLM query.
To evaluate the contribution of our seeding mechanism (Point 2 in \Cref{fig:llm_pipeline}), we also run Pynguin without seeding while fixing $t=60$. All other parameters are set to their default values, following past studies~\cite{codamosa}.
\Cref{tab:swtbench_p} presents the results; the last two columns report the number and percentage of \brts uniquely generated by the SBST Component, i.e., cases in which the LLM Component failed to produce an \brt.

With a 60-second time budget, Pynguin generates an F$\rightarrow$P test in 13/113 cases; however, our manual inspection revealed that only 10 are true \brts, while the others are similar to the case shown in \Cref{fig:f2p_but_not_br}.
Surprisingly, extending the time budget to 600 seconds results in one less F$\rightarrow$P test. Through manual analysis, we found that Pynguin initially generated tests for the focal method \texttt{nthroot\_mod()}, but after additional mutations, it discovered \texttt{sqrt\_mod()} and \texttt{primitive\_root()}, which achieved higher overall coverage, thus it discarded tests involving \texttt{nthroot\_mod()}.
This illustrates a key limitation of Pynguin: it optimizes for module-level coverage and does not focus on a specific focal method---a limitation we further discuss in~\Cref{ssec:case_studies}.

The last row of \Cref{tab:swtbench_p} reports the results of running Pynguin for 60 seconds without our seeding mechanism (Point 2 in~\Cref{fig:llm_pipeline}), thus initializing mutations from purely random tests. In this setting, only four F$\rightarrow$P tests are generated, highlighting the critical role of our seeding approach in aligning Pynguin's search with the issue description and the patch.

\smallskip
\noindent\textbf{Overall Contribution of SBST.}
One of our contributions is the introduction of SBST for generating \brts through a two-way orchestration between SBST and LLMs: SBST generates tests that are used as (i) candidate \brts (Point 9 in \Cref{fig:llm_pipeline}), and (ii) prompt context for an LLM that will generate candidate \brts (Point 5).
From \Cref{tab:swtbench_p} we see that Pynguin generates \brts for 6 entries where the LLM failed, and from \Cref{tab:res_rq1_ablation} we see that including the Pynguin-generated tests in the LLM prompt leads to 5 additional \brts, for a total of 11.
Hence, Pynguin contributes in 7.3\% (11/151) of the generated \brts of \ourtool.
In the following, we present two exemplary cases out of these eleven.

\begin{tcolorbox}
\begin{minipage}{\textwidth}
\textbf{Finding 2} \ourtool's LLM Component generates an \brt in 151/\N cases, with the focal context and the SBST tests being the most important prompt context.
\ourtool's SBST Component generates an \brt in 10 cases, 5 of which missed by the LLM Component.
Overall, SBST contributes to 11/151 (7.3\%) of the successful cases.
\end{minipage}
\end{tcolorbox}

\begin{figure}
    \centering
    \raggedright
    \includegraphics[trim=22pt 10pt 22pt 7pt, clip, width=\linewidth]{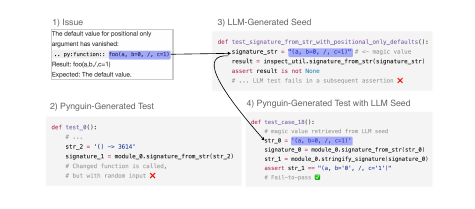}    \caption{Issue-reproducing test generated by Pynguin with gpt-4o seeds.}
    \label{fig:sbst_with_llm_seed example}
\end{figure}

\subsection{Analysis Of Success And Failure Cases}\label{ssec:case_studies}

We showcase the effective coordination between the two Components of \ourtool through two representative examples. 
Additionally, to examine \ourtool's limitations, we manually analyzed \num{50} cases in which it failed to generate an \brt.

\smallskip
\noindent\textbf{Representative Success Cases.}
Consider the issue of~\Cref{fig:sbst_with_llm_seed example}. For this issue, Pynguin without a seed generated the test shown in the bottom-left of the figure, where the patched function (\texttt{signature\_from\_str()}) is correctly called, but with a random input.
By starting the mutations from the LLM seed (top-right), the value required to reproduce the issue (highlighted in blue) is used instead of a random string, leading to the \brt shown in the bottom-right.

\begin{figure}
    \centering
    \includegraphics[trim=22pt 10pt 22pt 7pt, clip, width=\linewidth]{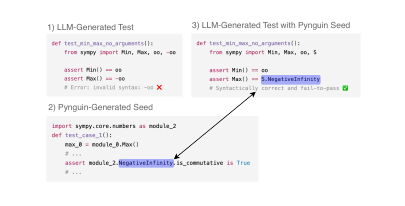}
    \caption{Issue-reproducing test generated by gpt-4o with Pynguin tests as prompt context.}
    \label{fig:llm_with_sbst_seed_example}
\end{figure}

Pynguin-generated tests can also serve as prompt context for the LLM, as we demonstrate with the issue of \Cref{fig:intro}, for which \gpt generated the test in the top-left of \Cref{fig:llm_with_sbst_seed_example}. 
This test throws an error instead of passing in the patched code, because the LLM considers \texttt{-oo} to be a singleton that, similar to \texttt{oo}, can be imported.
In reality, \texttt{-oo} is the result of the \texttt{\_\_neg\_\_} method of the \texttt{oo} singleton and as such cannot be imported.
For the same issue, Pynguin generated the test shown in the bottom-left of \Cref{fig:llm_with_sbst_seed_example}, which is not a F$\rightarrow$P test, but it uses the negative infinity properly, by importing \texttt{sympy.core.numbers.NegativeInfinity}.
When this syntactically correct test was included in the prompt, the LLM generated the \brt shown in the top-right of \Cref{fig:llm_with_sbst_seed_example}, which is almost the same as the top-left one but with \texttt{NegativeInfinity} instead of \texttt{-oo}.

\smallskip
\noindent\textbf{Analysis of Failure Cases.} %\label{sssec:case_studies_failure}
We manually analyze \num{30} cases where \ourtool's LLM Component failed to generate an \brt and 20 where the SBST Component failed and categorize the failure reasons using open card sorting~\cite{card_sorting}.

Regarding the LLM Component, the failure is due to mistakes in the LLM code in 22/30 cases, which we categorize in \textit{logical} (12), caused by an assertion error (or lack thereof), and \textit{syntactic} (10), where the LLM could not set up the test.
In only two of the syntactic failures further context could help, while for the other eight cases no context is missing: in two such examples, the LLM defines a helper function with two arguments and calls it later without arguments.

In five of the other eight cases, the LLM does \textit{not comply with requested output format}. 
For example, we ask for a raw test function and it returns a \href{https://docs.python.org/3/library/unittest.html#unittest.TestCase}{\texttt{unittest.TestCase}} class, which our injection is not designed to handle. 
This happens in Django issues because Django uses the \texttt{unittest} framework, so \gpt defaults to generating a unittest.TestCase despite instructed otherwise.
Finally, in three cases the issue is \textit{missing metadata}: the python version (the LLM code calls a deprecated method), installed pip packages (the LLM imports a non-installed package), and the exact date when the test is run (the test must use a future date, and the LLM uses 2023, which is not the future anymore).

Regarding the SBST Component, the most common issue (10/20) is that the tests do not cover the changes, because Pynguin generates tests for whole modules and not classes or methods.
In 5/20 cases, the test requires setup currently unsupported by Pynguin, e.g., the patch only adds decorators to methods, and Pynguin does not support decorators.
In 3/20 cases, the changes are covered but not triggered, while in 2/20 cases Pynguin generates tests that require additional setup to run, leading to errors.
We release the analyzed examples along with our observations in our replication package~\cite{replication_package}.

\section{In Vivo Evaluation}\label{sec:evaluation_online}
Despite increasing research on the automated generation of \brts, the evaluation of these techniques in real-world development environments is lacking.
Such evaluations are crucial for at least two reasons: First, existing benchmarks---including those based on SWE-Bench~\cite{tddbench,swtbench}---rely on historical data, in some cases dating back to 2013, thus raising risks of LLM memorization~\cite{memorization}.
Second, the ultimate goal of automation is to assist developers; therefore, developer perception and output usability are to be taken into account.
Ignoring this dimension risks painting a partial picture, as illustrated in the domain of reviewer recommendation: although some models achieved up to 92\% accuracy on historical benchmarks~\cite{reviewer_recom}, these models proved not useful when deployed in real development settings~\cite{review_recommender}.

We conduct a real-world evaluation of \ourtool by deploying it in three open source repositories for three months, allowing us to both evaluate \ourtool in a memorization-free setting and collect developers' perceptions on the generated tests.

\subsection{Evaluation Methodology}\label{ssec:evaluation_meth_invivo}
We develop and open-source a GitHub bot that, once deployed in a repository, is triggered when a PR is opened.
Then, if the PR (i) resolves an issue, (ii) edits at least one \texttt{.py} file, and (iii) does not accompany the patch with a test, the bot retrieves the patch and the issue and feeds them into \ourtool.
If \ourtool generates a F$\rightarrow$P test, our bot proposes it to the developer as a PR comment.
This allows us to evaluate \ourtool's performance in a memorization-free setting, collect developers' perceptions of the tests, and surface challenges for researchers and practitioners based on developers' feedback. We focus on the following research questions:

\begin{tcolorbox}
\begin{minipage}{\textwidth}
\textbf{RQ3.1.} How does \ourtool perform  in a real-world setting? \textbf{RQ3.2.} How do developers perceive the \brts it generates?
\end{minipage}
\end{tcolorbox}

\begin{table}
\centering
\scriptsize
\caption{Mozilla repositories used in our study.}
\label{tab:moz_repos}
\begin{tabular}{llrrr}
\textbf{Repository} & \textbf{Description} & \textbf{\# stars} & \textbf{\# forks} & \textbf{\# PRs} \\
\midrule
bugbug & Platform for ML on SE  & 537 & 311 & 26  \\
bugbot & Mozilla release management tool & 48  & 71 & 5 \\
libmozdata & Library to access Mozilla data & 12 & 13  & 1\\
\end{tabular}
\end{table}

We deployed the bot for three months across three open-source repositories within the Mozilla corporation, as detailed in \Cref{tab:moz_repos}, collecting data from \num{32} pull requests (PRs) in total.
For each PR, we executed the LLM Component of \ourtool once for each of the three LLMs of~\Cref{tab:results_rq1_autotdd}, to compare their performance and maximize the number of tests eligible for developer feedback.
To avoid overwhelming contributors---which prior work suggests may reduce the quality of feedback~\cite{survey_fatigue}---we proposed only a single test per PR.
When both Components of \ourtool produced an \brt for the same PR, we prioritized the LLM-generated test, due to previous work suggesting that SBST tests are often rejected by developers due to reduced readability~\cite{sbst_readability}.
In cases where multiple LLMs generated an \brt, we selected the test from the higher-performing LLM in~\Cref{tab:results_rq1_autotdd}.
All the F$\rightarrow$P tests for a given PR that were not proposed to the developer, were manually assessed by the first author to determine whether they are \brts.

For each proposed test, we ask the maintainer(s) of the repos to provide feedback on the test in two steps.
First, we ask if they consider the proposed F$\rightarrow$P test to be an \brt, followed by an open answer to the following question: ``\textit{Reason for not accepting the test, or minor changes you would require to accept it, or general comment if you will accept it as is.}''
We give the option to request minor changes, because previous work has shown that when proposing automatically generated tests to open-source maintainers, they often request such changes~\cite{shaken_stirred}.

In total, we propose tests in 11 out of \num{32} PRs, each represented as a column in \Cref{tab:results_moz_all_prs}, where the rows represent the two Components of \ourtool.
The F$\rightarrow$P tests that the developer judged to be issue-reproducing are denoted with \ding{51}, while the rest are marked with \ding{109}.
The \brts that were also merged in the project are shown with \textcolor{ForestGreen}{\ding{51}}.
Finally, PRs for which Pynguin could not run successfully are denoted with \ding{55}.
We discuss the developer feedback in the next section.

\subsection{Results}\label{ssec:results_invivo}
\begin{table}
\centering
\scriptsize
\caption{PRs for which at least a F$\rightarrow$P test was generated. \ding{51} denotes an \brt; \ding{109} denotes a F$\rightarrow$P that is not an \brt;  \textcolor{ForestGreen}{\ding{51}} denotes a developer-accepted test; \ding{55} denotes that Pynguin could not run in this PR.}\label{tab:results_moz_all_prs}
\begin{tabular}{l|CCCCCCCCCCC}

\textbf{Comp.$\downarrow$ / PR  $\rightarrow$} & 1 & 2 & 3 & 4 & 5 & 6 & 7 & 8 & 9 & 10 & 11 \\
\midrule
    \hspace{0.74cm} gpt-4o     & \textcolor{ForestGreen}{\ding{51}} & \ding{51} & \ding{51} &   &   &   \textcolor{ForestGreen}{\ding{51}} & &  \ding{51} & \ding{51} &  &  \\
LLM \hspace{0.15cm} llama      &   &   & \ding{51} & \ding{109} &   & \ding{51} & \ding{109} & \ding{51} & \ding{51} & \ding{109} & \ding{109} \\
    \hspace{0.74cm} DeepSeek   &   &   & \ding{51} & \ding{109} & \ding{109} &   &   & \ding{51} &   &   & \ding{109} \\
\midrule
Pynguin                        & \ding{55} & \ding{55} & \ding{55} & \ding{55}  &  &  &   & \ding{109} & \ding{109} & \ding{55} &  \\
\end{tabular}
\end{table}

In contrast to our benchmark-based evaluation, \llama achieves the highest number of F$\rightarrow$P tests (8/32, 25.0\%), followed by \gpt (7/32, 21.9\%) and \deepseek (4/32, 12.5\%).
The smaller number of parameters of \llama could lead to less memorization than larger models, and hence, lower performance on historical data.
Gpt-4o has been shown to memorize more than \llama in non-SE tasks~\cite{llm_memorization}, however, a larger sample is needed to study this difference in more detail.

Pynguin successfully ran in 11/32 PRs and generated two F$\rightarrow$P tests, but our bot did not propose them to the developer because \gpt-generated tests had precedence.
After inspecting the tests ourselves, we found that they were not \brts.
The 11 PRs where Pynguin ran successfully is a small sample, and since SBST tools do not suffer from memorization, Pynguin's performance on the historical data is expected to be the same as in new data.

The above suggest that \ourtool can generate \brts in a real-world setting, beyond benchmarks of historical, potentially memorized data.
The accuracy of \ourtool in the real-world setting is lower than in TDD-Bench-Verified, but we cannot attribute the drop to memorization only, since the repositories are also different.

\begin{tcolorbox}
\begin{minipage}{\textwidth}
\textbf{Finding 3.1.} \ourtool generated fail-to-pass tests in 8/32 (25\%) PRs when deployed in three open-source repos, with \llama outperforming larger models.
\end{minipage}
\end{tcolorbox}

We analyze here the developer feedback for the 11 proposed tests.
The developer considered 6/11 (55\%) tests as valid \brts, two of which were successfully included in the projects' test suite, one without any changes and the other after inserting it in a different test file.
In one case the test was not added because the PR was closed without merging, while in another case, the developer said that the test reproduces the issue, but ``it does not test a critical part,'' i.e., the issue was not critical enough to require a test.
Finally, in two cases the developer deemed the test to be reproducing the issue, but with excessive mocking that missed important aspects of the implementation. Specifically, the PR patched a crash caused by the field \texttt{comments} missing from a data structure.
To patch this bug, the developer created the function \texttt{handle\_missing\_comments()} to add the field if missing.
The proposed test set up the data with the missing field, but simply asserted that \texttt{handle\_missing\_comments()} was called, while it should also check that the field exists.

We analyze below the 5/11 PRs in which the developer judged the F$\rightarrow$P test as \emph{not} issue-reproducing.
In two PRs, the issue was a renaming of a component from \texttt{Fenix} to \texttt{Firefox for Android};
\ourtool generated a test that asserts the new name, but for such patches, an accompanying test was not necessary.
In two other cases, the issue was requesting a new feature and the maintainer explained: ``The test would be perfect if this were a regression, but it is a feature.''
Upon a follow-up question, the maintainer clarified that tests for new features are generally useful, but for these specific two they were not that useful.
Finally, the last PR resolves a bug where a query incorrectly ignored some entries, by changing the query parameters.
In the lack of context for a proper \brt, \ourtool asserts that the query parameters have changed, which is not reproducing the issue.

The above feedback highlights key distinctions between SWE-Bench-based evaluations and real-world deployments.
In \swebench, a relevant test is guaranteed to exist for each patch because \swebench consists of patches that i) resolve an issue, \textit{and} ii) include a developer-written test. 
In contrast, our evaluation focuses on the utility of automatically generated tests \emph{in the absence of developer-written ones}, which also resulted in the bot proposing tests even when no tests were relevant for the patch.
To mitigate these false positives, one developer suggested limiting bot activation to issues labeled as ``bug'' or allowing developers to invoke the bot on demand.

Another developer explained that the proposed tests become obsolete when new commits are pushed to the PR, and suggested triggering the bot on each new commit, which we implemented.
We note, however, that for multi-revision PRs, this could swamp the comment section.
An on-demand bot could also help in this case.

\begin{tcolorbox}
\begin{minipage}{\textwidth}
\textbf{Finding 3.2.} Of the 11 proposed tests, 6 were considered valid \brts and 2 were integrated in the test suite. 
Reasons for not considering the tests are that, in contrast to benchmarks, some issues do not require an accompanying test and that the test uses excessive mocking, missing core functionality.
\end{minipage}
\end{tcolorbox}

\section{Discussion}\label{sec:discussion}
In this section, we discuss the implications of our findings for future research and practice, along with the limitations of \ourtool and the threats to the validity of our study.

\smallskip
\noindent\textbf{On Benchmarks and Metrics.}
We underline the importance of manually inspecting benchmarks, as our analysis of \tddbv revealed \num{23} trivial issues that could inflate performance.
Furthermore, as a community we should reflect on evaluation metrics like F$\rightarrow$P, which, while widely used, do not always correspond to \brts.
Manual validation, developer validation, and matching the failure trace to the issue, as done by Issue2Test~\cite{issue2test}, are possible ways forward.

\smallskip
\noindent\textbf{Recommendations for Practitioners.}
For developers deploying test generation tools, we recommend being aware of cases where a patch does not require a test and syncing with developers to agree upon the tool \emph{triggering criteria}.
PRs often undergo many revisions, thus generating one test per revision may overwhelm the developers.
Other options like generating whenever a PR is marked as ``ready'' could be beneficial. On-demand tools may address both challenges, though they shift some responsibility to developers, potentially reducing automation.

\smallskip
\noindent\textbf{Recommendations for Researchers.}
While improving upon the SOTA, \ourtool manages to generate an \brt for \num{35.4}\% of the issue-patch pairs in the benchmark and for \num{25.0}\% in the real-world evaluation, which leaves ample room for future improvement.
We hope that our manual analysis of 50 issues where \ourtool failed can provide directions for future research.

\ourtool provides evidence that going beyond only LLMs for test generation can be beneficial, as indicated by the contribution of SBST in \num{7.3}\% of the \brts.
However, the contribution of SBST is relatively low, which we trace back to two main \textit{limitations}. 
The first limitation is 
the inability of SBST tools to run in arbitrary code, e.g., Pynguin runs successfully in 113/426 (26\%) dataset instances. If future Pynguin versions or new SBST tools add support for more complex code, the contribution of Pynguin-generated tests would subsequently increase.
The second limitation is the inability of SBST tools to generate tests for specific methods or lines instead of whole modules, which leads to SBST-generated tests not covering the patch in 10/20 failure cases manually analyzed in~\Cref{ssec:results_invivo}.
More advanced SBST tools that support test generation for fine-grained targets like functions or lines would mitigate this limitation and further improve the contribution of SBST.

\smallskip
\noindent\textbf{Threats to Validity.}
Our study includes two manual inspection steps: the filtering of \tddbv for trivial entries, and the judgment of whether a F$\rightarrow$P test is an \brt, and as such these inspections could be prone to human error.
We release the inspected entries to be independently validated.

Our experiments focus on Python and use SWE-Bench-derived benchmarks, hence our results may not generalize to other languages or codebases.
The real-world study includes \num{32} PRs, which limits generalization, but provides a solid first step towards evaluating similar tools with developers.

Both LLMs and Pynguin are non-deterministic systems. 
To boost reproducibility, we used a fixed seed for Pynguin, but running on different hardware could result in slightly different results.
To mitigate the non-determinism of LLMs we used a temperature of zero and pinned the versions of each model.
However, repeating each experiment multiple times, with different seeds for Pynguin, would have further strengthened the statistical power of our experiments, and the absence of such repetitions is a potential threat to validity.

The authors of the two baselines could not provide the replication package at the moment, so we implemented the approaches based on their paper.
Our implementation closely matches the reported performance (23.5\% vs 24.3\%) using the same pinned version of \gpt. 

\section{Conclusion}\label{sec:conclusion}
In this paper, we presented \ourtool, a hybrid approach that combines LLMs and SBST to generate \brts. 
By statically retrieving and generating context for the LLM and adapting SBST for generating \brts, \ourtool outperforms the state-of-the-art, achieving \bestpct success rate. 
We evaluated \ourtool in a widely used benchmark of historical data to understand how different design decisions affect its performance.
By further evaluating \ourtool in vivo, we increased our confidence about its ability to perform in memorization-free environments, and also gathered developer feedback that could be helpful for future research and practice.

\section*{Acknowledgments}
K. Kitsios and A. Bacchelli gratefully acknowledge the support of the Swiss National Science
Foundation through the SNF Project 200021\_197227.

\bibliographystyle{IEEEtranN}
\bibliography{references}

% Generated by IEEEtranN.bst, version: 1.14 (2015/08/26)
\begin{thebibliography}{51}
\providecommand{\natexlab}[1]{#1}
\providecommand{\url}[1]{#1}
\csname url@samestyle\endcsname
\providecommand{\newblock}{\relax}
\providecommand{\bibinfo}[2]{#2}
\providecommand{\BIBentrySTDinterwordspacing}{\spaceskip=0pt\relax}
\providecommand{\BIBentryALTinterwordstretchfactor}{4}
\providecommand{\BIBentryALTinterwordspacing}{\spaceskip=\fontdimen2\font plus
\BIBentryALTinterwordstretchfactor\fontdimen3\font minus \fontdimen4\font\relax}
\providecommand{\BIBforeignlanguage}[2]{{%
\expandafter\ifx\csname l@#1\endcsname\relax
\typeout{** WARNING: IEEEtranN.bst: No hyphenation pattern has been}%
\typeout{** loaded for the language `#1'. Using the pattern for}%
\typeout{** the default language instead.}%
\else
\language=\csname l@#1\endcsname
\fi
#2}}
\providecommand{\BIBdecl}{\relax}
\BIBdecl

\bibitem[Bettenburg et~al.(2008)Bettenburg, Just, Schr{\"o}ter, Weiss, Premraj, and Zimmermann]{bettenburg2008makes}
N.~Bettenburg, S.~Just, A.~Schr{\"o}ter, C.~Weiss, R.~Premraj, and T.~Zimmermann, ``What makes a good bug report?'' in \emph{Proceedings of the 16th ACM SIGSOFT International Symposium on Foundations of Software Engineering}, 2008, pp. 308--318.

\bibitem[Kang et~al.(2023)Kang, Yoon, and Yoo]{libro}
S.~Kang, J.~Yoon, and S.~Yoo, ``Large language models are few-shot testers: Exploring llm-based general bug reproduction,'' in \emph{2023 IEEE/ACM 45th International Conference on Software Engineering (ICSE)}.\hskip 1em plus 0.5em minus 0.4em\relax IEEE, 2023, pp. 2312--2323.

\bibitem[Kochhar et~al.(2019)Kochhar, Xia, and Lo]{developer_survey_on_regr_tests}
P.~S. Kochhar, X.~Xia, and D.~Lo, ``Practitioners' views on good software testing practices,'' in \emph{2019 IEEE/ACM 41st International Conference on Software Engineering: Software Engineering in Practice (ICSE-SEIP)}.\hskip 1em plus 0.5em minus 0.4em\relax IEEE, 2019, pp. 61--70.

\bibitem[Wong et~al.(1997)Wong, Horgan, London, and Agrawal]{motivation1}
W.~E. Wong, J.~R. Horgan, S.~London, and H.~Agrawal, ``A study of effective regression testing in practice,'' in \emph{PROCEEDINGS The Eighth International Symposium On Software Reliability Engineering}.\hskip 1em plus 0.5em minus 0.4em\relax IEEE, 1997, pp. 264--274.

\bibitem[Onoma et~al.(1998)Onoma, Tsai, Poonawala, and Suganuma]{motivation2}
A.~K. Onoma, W.-T. Tsai, M.~Poonawala, and H.~Suganuma, ``Regression testing in an industrial environment,'' \emph{Communications of the ACM}, vol.~41, no.~5, pp. 81--86, 1998.

\bibitem[Wang et~al.(2023)Wang, Lian, Marinov, and Xu]{motivation3}
S.~Wang, X.~Lian, D.~Marinov, and T.~Xu, ``Test selection for unified regression testing,'' in \emph{2023 IEEE/ACM 45th International Conference on Software Engineering (ICSE)}.\hskip 1em plus 0.5em minus 0.4em\relax IEEE, 2023, pp. 1687--1699.

\bibitem[Likert(1932)]{likertscale}
R.~Likert, ``A technique for the measurement of attitudes.'' \emph{Archives of psychology}, 1932.

\bibitem[Labuschagne et~al.(2017)Labuschagne, Inozemtseva, and Holmes]{timeconsuming1}
A.~Labuschagne, L.~Inozemtseva, and R.~Holmes, ``Measuring the cost of regression testing in practice: A study of java projects using continuous integration,'' in \emph{Proceedings of the 2017 11th joint meeting on foundations of software engineering}, 2017, pp. 821--830.

\bibitem[Straubinger and Fraser(2023)]{timeconsuming2}
P.~Straubinger and G.~Fraser, ``A survey on what developers think about testing,'' in \emph{2023 IEEE 34th International Symposium on Software Reliability Engineering (ISSRE)}.\hskip 1em plus 0.5em minus 0.4em\relax IEEE, 2023, pp. 80--90.

\bibitem[Levin and Yehudai(2017)]{regression_tests_ignored}
S.~Levin and A.~Yehudai, ``The co-evolution of test maintenance and code maintenance through the lens of fine-grained semantic changes,'' in \emph{2017 IEEE International Conference on Software Maintenance and Evolution (ICSME)}, 2017, pp. 35--46.

\bibitem[Ahmed et~al.(2024)Ahmed, Hirzel, Pan, Shinnar, and Sinha]{tddbench}
T.~Ahmed, M.~Hirzel, R.~Pan, A.~Shinnar, and S.~Sinha, ``Tdd-bench verified: Can llms generate tests for issues before they get resolved?'' \emph{arXiv preprint arXiv:2412.02883}, 2024.

\bibitem[M{\"u}ndler et~al.(2024)M{\"u}ndler, M{\"u}ller, He, and Vechev]{swtbench}
N.~M{\"u}ndler, M.~M{\"u}ller, J.~He, and M.~Vechev, ``Swt-bench: Testing and validating real-world bug-fixes with code agents,'' \emph{Advances in Neural Information Processing Systems}, vol.~37, pp. 81\,857--81\,887, 2024.

\bibitem[Nashid et~al.(2025)Nashid, Bouzenia, Pradel, and Mesbah]{issue2test}
N.~Nashid, I.~Bouzenia, M.~Pradel, and A.~Mesbah, ``Issue2test: Generating reproducing test cases from issue reports,'' \emph{arXiv preprint arXiv:2503.16320}, 2025.

\bibitem[Ahmed et~al.(2025)Ahmed, Ganhotra, Pan, Shinnar, Sinha, and Hirzel]{otter}
\BIBentryALTinterwordspacing
T.~Ahmed, J.~Ganhotra, R.~Pan, A.~Shinnar, S.~Sinha, and M.~Hirzel, ``Otter: Generating tests from issues to validate swe patches,'' \emph{arXiv}, 2025. [Online]. Available: \url{https://arxiv.org/abs/2502.05368}
\BIBentrySTDinterwordspacing

\bibitem[Liu et~al.(2025)Liu, Lee, Losiouk, and B{\"o}hme]{fuzzing_regression_tests}
J.~Liu, S.~Lee, E.~Losiouk, and M.~B{\"o}hme, ``Can llm generate regression tests for software commits?'' \emph{arXiv preprint arXiv:2501.11086}, 2025.

\bibitem[Eghbali and Pradel(2024)]{hallucinations}
A.~Eghbali and M.~Pradel, ``De-hallucinator: Mitigating llm hallucinations in code generation tasks via iterative grounding,'' \emph{arXiv preprint arXiv:2401.01701}, 2024.

\bibitem[Kitsios et~al.(2025)Kitsios, Castelluccio, and Bacchelli]{replication_package}
\BIBentryALTinterwordspacing
K.~Kitsios, M.~Castelluccio, and A.~Bacchelli. (2025, Aug.) Replication package for "automated generation of issue-reproducing tests by combining llms and search-based testing". [Online]. Available: \url{https://doi.org/10.5281/zenodo.16949043}
\BIBentrySTDinterwordspacing

\bibitem[Harman et~al.(2012)Harman, Mansouri, and Zhang]{harman2012search}
M.~Harman, S.~A. Mansouri, and Y.~Zhang, ``Search-based software engineering: Trends, techniques and applications,'' \emph{ACM Computing Surveys (CSUR)}, vol.~45, no.~1, pp. 1--61, 2012.

\bibitem[Hooda et~al.(2024)Hooda, Christodorescu, Allamanis, Wilson, Fawaz, and Jha]{memorization}
A.~Hooda, M.~Christodorescu, M.~Allamanis, A.~Wilson, K.~Fawaz, and S.~Jha, ``Do large code models understand programming concepts? counterfactual analysis for code predicates,'' in \emph{Forty-first International Conference on Machine Learning}, 2024.

\bibitem[Cohen-Inger et~al.(2025)Cohen-Inger, Elisha, Shapira, Rokach, and Cohen]{llm_memorization}
\BIBentryALTinterwordspacing
N.~Cohen-Inger, Y.~Elisha, B.~Shapira, L.~Rokach, and S.~Cohen, ``Forget what you know about llms evaluations -- llms are like a chameleon,'' 2025. [Online]. Available: \url{https://arxiv.org/abs/2502.07445}
\BIBentrySTDinterwordspacing

\bibitem[Kovalenko et~al.(2018)Kovalenko, Tintarev, Pasynkov, Bird, and Bacchelli]{review_recommender}
V.~Kovalenko, N.~Tintarev, E.~Pasynkov, C.~Bird, and A.~Bacchelli, ``Does reviewer recommendation help developers?'' \emph{IEEE Transactions on Software Engineering}, vol.~46, no.~7, pp. 710--731, 2018.

\bibitem[Vaithilingam et~al.(2022)Vaithilingam, Zhang, and Glassman]{code_completion}
P.~Vaithilingam, T.~Zhang, and E.~L. Glassman, ``Expectation vs. experience: Evaluating the usability of code generation tools powered by large language models,'' in \emph{Chi conference on human factors in computing systems extended abstracts}, 2022, pp. 1--7.

\bibitem[Balachandran(2013)]{reviewer_recom}
V.~Balachandran, ``Reducing human effort and improving quality in peer code reviews using automatic static analysis and reviewer recommendation,'' in \emph{2013 35th International Conference on Software Engineering (ICSE)}.\hskip 1em plus 0.5em minus 0.4em\relax IEEE, 2013, pp. 931--940.

\bibitem[Lukasczyk and Fraser(2022)]{pynguin}
S.~Lukasczyk and G.~Fraser, ``Pynguin: Automated unit test generation for python,'' in \emph{Proceedings of the ACM/IEEE 44th International Conference on Software Engineering: Companion Proceedings}, 2022, pp. 168--172.

\bibitem[Lukasczyk et~al.(2023)Lukasczyk, Kroi{\ss}, and Fraser]{pynguin_time_budget}
S.~Lukasczyk, F.~Kroi{\ss}, and G.~Fraser, ``An empirical study of automated unit test generation for python,'' \emph{Empirical Software Engineering}, vol.~28, no.~2, p.~36, 2023.

\bibitem[Lemieux et~al.(2023)Lemieux, Inala, Lahiri, and Sen]{codamosa}
C.~Lemieux, J.~P. Inala, S.~K. Lahiri, and S.~Sen, ``Codamosa: Escaping coverage plateaus in test generation with pre-trained large language models,'' in \emph{2023 IEEE/ACM 45th International Conference on Software Engineering (ICSE)}.\hskip 1em plus 0.5em minus 0.4em\relax IEEE, 2023, pp. 919--931.

\bibitem[Ryan et~al.(2024)Ryan, Jain, Shang, Wang, Ma, Ramanathan, and Ray]{testgen_llm1}
G.~Ryan, S.~Jain, M.~Shang, S.~Wang, X.~Ma, M.~K. Ramanathan, and B.~Ray, ``Code-aware prompting: A study of coverage-guided test generation in regression setting using llm,'' \emph{Proceedings of the ACM on Software Engineering}, vol.~1, no. FSE, pp. 951--971, 2024.

\bibitem[Fraser and Arcuri(2011)]{evosuite}
G.~Fraser and A.~Arcuri, ``Evosuite: automatic test suite generation for object-oriented software,'' in \emph{Proceedings of the 19th ACM SIGSOFT symposium and the 13th European conference on Foundations of software engineering}, 2011, pp. 416--419.

\bibitem[Holland(1992)]{genetic_algorithms}
J.~H. Holland, \emph{Adaptation in natural and artificial systems: an introductory analysis with applications to biology, control, and artificial intelligence}.\hskip 1em plus 0.5em minus 0.4em\relax MIT press, 1992.

\bibitem[Kirkpatrick et~al.(1983)Kirkpatrick, Gelatt~Jr, and Vecchi]{simulated_annealing}
S.~Kirkpatrick, C.~D. Gelatt~Jr, and M.~P. Vecchi, ``Optimization by simulated annealing,'' \emph{science}, vol. 220, no. 4598, pp. 671--680, 1983.

\bibitem[Xie(2006)]{regression_assertion}
T.~Xie, ``Augmenting automatically generated unit-test suites with regression oracle checking,'' in \emph{European Conference on Object-Oriented Programming}.\hskip 1em plus 0.5em minus 0.4em\relax Springer, 2006, pp. 380--403.

\bibitem[Guilherme and Vincenzi(2023)]{testgen_llm2}
V.~Guilherme and A.~Vincenzi, ``An initial investigation of chatgpt unit test generation capability,'' in \emph{Proceedings of the 8th Brazilian Symposium on Systematic and Automated Software Testing}, 2023, pp. 15--24.

\bibitem[Zhang et~al.(2024)Zhang, Ruan, Fan, and Roychoudhury]{autocoderover}
Y.~Zhang, H.~Ruan, Z.~Fan, and A.~Roychoudhury, ``Autocoderover: Autonomous program improvement,'' in \emph{Proceedings of the 33rd ACM SIGSOFT International Symposium on Software Testing and Analysis}, 2024, pp. 1592--1604.

\bibitem[Bouzenia et~al.(2024)Bouzenia, Devanbu, and Pradel]{repairagent}
I.~Bouzenia, P.~Devanbu, and M.~Pradel, ``Repairagent: an autonomous, llm-based agent for program repair.(2024),'' \emph{arXiv preprint arXiv:2403.17134}, 2024.

\bibitem[Shamshiri(2015)]{evosuiter}
S.~Shamshiri, ``Automated unit test generation for evolving software,'' in \emph{Proceedings of the 2015 10th Joint Meeting on Foundations of Software Engineering}, 2015, pp. 1038--1041.

\bibitem[McKeeman(1998)]{differential_testing}
W.~M. McKeeman, ``Differential testing for software,'' \emph{Digital Technical Journal}, vol.~10, no.~1, pp. 100--107, 1998.

\bibitem[Grano et~al.(2018)Grano, Scalabrino, Gall, and Oliveto]{sbst_readability}
G.~Grano, S.~Scalabrino, H.~C. Gall, and R.~Oliveto, ``An empirical investigation on the readability of manual and generated test cases,'' in \emph{Proceedings of the 26th Conference on Program Comprehension}, 2018, pp. 348--351.

\bibitem[Just et~al.(2014)Just, Jalali, and Ernst]{defects4j}
R.~Just, D.~Jalali, and M.~D. Ernst, ``Defects4j: A database of existing faults to enable controlled testing studies for java programs,'' in \emph{Proceedings of the 2014 international symposium on software testing and analysis}, 2014, pp. 437--440.

\bibitem[Causevic et~al.(2011)Causevic, Sundmark, and Punnekkat]{tdd_in_practice}
A.~Causevic, D.~Sundmark, and S.~Punnekkat, ``Factors limiting industrial adoption of test driven development: A systematic review,'' in \emph{2011 Fourth IEEE International Conference on Software Testing, Verification and Validation}.\hskip 1em plus 0.5em minus 0.4em\relax IEEE, 2011, pp. 337--346.

\bibitem[Fioraldi et~al.(2020)Fioraldi, Maier, Ei{\ss}feldt, and Heuse]{aflpp}
A.~Fioraldi, D.~Maier, H.~Ei{\ss}feldt, and M.~Heuse, ``$\{$AFL++$\}$: Combining incremental steps of fuzzing research,'' in \emph{14th USENIX workshop on offensive technologies (WOOT 20)}, 2020.

\bibitem[Gruber et~al.(2024)Gruber, Roslan, Parry, Scharnb{\"o}ck, McMinn, and Fraser]{sbst_flaky}
M.~Gruber, M.~F. Roslan, O.~Parry, F.~Scharnb{\"o}ck, P.~McMinn, and G.~Fraser, ``Do automatic test generation tools generate flaky tests?'' in \emph{Proceedings of the 46th IEEE/ACM International Conference on Software Engineering}, 2024, pp. 1--12.

\bibitem[Gote et~al.(2021)Gote, Scholtes, and Schweitzer]{coupling1}
C.~Gote, I.~Scholtes, and F.~Schweitzer, ``Analysing time-stamped co-editing networks in software development teams using git2net,'' \emph{Empirical software engineering}, vol.~26, no.~4, p.~75, 2021.

\bibitem[Gall et~al.(1998)Gall, Hajek, and Jazayeri]{coupling2}
H.~Gall, K.~Hajek, and M.~Jazayeri, ``Detection of logical coupling based on product release history,'' in \emph{Proceedings. International Conference on Software Maintenance (Cat. No. 98CB36272)}.\hskip 1em plus 0.5em minus 0.4em\relax IEEE, 1998, pp. 190--198.

\bibitem[Tufano et~al.(2020)Tufano, Drain, Svyatkovskiy, Deng, and Sundaresan]{focal_context}
M.~Tufano, D.~Drain, A.~Svyatkovskiy, S.~K. Deng, and N.~Sundaresan, ``Unit test case generation with transformers and focal context,'' \emph{arXiv preprint arXiv:2009.05617}, 2020.

\bibitem[Ramos et~al.(2024)Ramos, Mamede, Jain, Canelas, Gamboa, and Goues]{defects4j_saturated}
D.~Ramos, C.~Mamede, K.~Jain, P.~Canelas, C.~Gamboa, and C.~L. Goues, ``Are large language models memorizing bug benchmarks?'' \emph{arXiv preprint arXiv:2411.13323}, 2024.

\bibitem[Jimenez et~al.(2024)Jimenez, Yang, Wettig, Yao, Pei, Press, and Narasimhan]{swebench}
\BIBentryALTinterwordspacing
C.~E. Jimenez, J.~Yang, A.~Wettig, S.~Yao, K.~Pei, O.~Press, and K.~R. Narasimhan, ``{SWE}-bench: Can language models resolve real-world github issues?'' in \emph{The Twelfth International Conference on Learning Representations}, 2024. [Online]. Available: \url{https://openreview.net/forum?id=VTF8yNQM66}
\BIBentrySTDinterwordspacing

\bibitem[Hilton et~al.(2018)Hilton, Bell, and Marinov]{patch_cov}
M.~Hilton, J.~Bell, and D.~Marinov, ``A large-scale study of test coverage evolution,'' in \emph{Proceedings of the 33rd ACM/IEEE International Conference on Automated Software Engineering}, 2018, pp. 53--63.

\bibitem[Peysakhovich and Lerer(2023)]{recency_bias}
A.~Peysakhovich and A.~Lerer, ``Attention sorting combats recency bias in long context language models,'' \emph{arXiv preprint arXiv:2310.01427}, 2023.

\bibitem[Spencer(2009)]{card_sorting}
D.~Spencer, \emph{Card sorting: Designing usable categories}.\hskip 1em plus 0.5em minus 0.4em\relax Rosenfeld Media, 2009.

\bibitem[Fass-Holmes(2022)]{survey_fatigue}
B.~Fass-Holmes, ``Survey fatigue--what is its role in undergraduates' survey participation and response rates?.'' \emph{Journal of interdisciplinary Studies in Education}, vol.~11, no.~1, pp. 56--73, 2022.

\bibitem[Brandt et~al.(2024)Brandt, Khatami, Wessel, and Zaidman]{shaken_stirred}
C.~Brandt, A.~Khatami, M.~Wessel, and A.~Zaidman, ``Shaken, not stirred. how developers like their amplified tests,'' \emph{IEEE Transactions on Software Engineering}, 2024.

\end{thebibliography}

\end{document}